\documentclass[traditabstract]{aa}
\usepackage{multirow}
\usepackage{graphics}
\usepackage{txfonts}
\usepackage{amssymb}
\usepackage{lscape}
\usepackage{epsfig}
\usepackage{natbib}
\begin{document}

\def\mpc{h^{-1} {\rm{Mpc}}} 
\def\kpc{h^{-1} {\rm{kpc}}}
\newcommand{\mincir}{\raise
-2.truept\hbox{\rlap{\hbox{$\sim$}}\raise5.truept\hbox{$<$}\ }}
\newcommand{\magcir}{\raise
-2.truept\hbox{\rlap{\hbox{$\sim$}}\raise5.truept\hbox{$>$}\ }}

\title{The XXL Survey\thanks{Based on observations obtained with {\it XMM-Newton}, an ESA science 
mission with instruments and contributions directly funded by
ESA Member States and NASA.}} 
\subtitle{XIX. A realistic population of simulated X-ray AGN:\\ Comparison of models with observations}

\author{E. Koulouridis\inst{1,2} \and L. Faccioli\inst{1,2} \and A. M. C. Le Brun\inst{1,2} \and M. Plionis\inst{3,4} \and I. G. McCarthy\inst{5} \and M. Pierre\inst{1,2}
\and A. Akylas\inst{6} \and I. Georgantopoulos\inst{6} \and S. Paltani \inst{7} \and C. Lidman\inst{8} \and S. Fotopoulou\inst{7} \and C. Vignali\inst{9,10} \and F. Pacaud\inst{11} \and P. Ranalli\inst{12}}

\institute{IRFU, CEA, Universit\'e Paris-Saclay, F-91191 Gif-sur-Yvette, France
\and Universit\'e Paris Diderot, AIM, Sorbonne Paris Cit\'e, CEA, CNRS, F-91191
Gif-sur-Yvette, France
\and National Observatory of Athens, Lofos Nymfon, Thessio, 11851 Athens, Greece
\and Physics Department of Aristotle University of Thessaloniki, University Campus, 54124, Thessaloniki, Greece
\and Astrophysics Research Institute, Liverpool John Moores University, 146 Brownlow Hill, Liverpool L3 5RF
\and Institute for Astronomy \& Astrophysics, Space Applications \& Remote Sensing, National Observatory of Athens, Palaia Penteli 15236, Athens, Greece
\and Department of Astronomy, University of Geneva, ch. d'Ecogia 16, 1290, Versoix, Switzerland
\and Australian Astronomical Observatory, North Ryde, NSW 2113, Australia
\and Dipartimento di Fisica e Astronomia, Alma Mater Studiorum, Universit\`a degli Studi di Bologna, Via Gobetti 93/2, 40129 Bologna, Italy
\and INAF -- Osservatorio Astronomico di Bologna, Via Gobetti 93/3, 40129 Bologna, Italy
\and Argelander-Institut f\"ur Astronomie, University of Bonn, Auf dem H\"ugel 71, 53121 Bonn, Germany
\and Lund Observatory, Box 43, 22100 Lund, Sweden}

\date{Received/Accepted}

\abstract{Modern cosmological simulations rely heavily on feedback from active galactic nuclei (AGN) in order to stave off overcooling in massive galaxies and galaxy groups and clusters. Given that AGN are a key component of such simulations, an important independent test is whether or not the simulations capture the broad demographics of the observed AGN population. However, to date, comparisons between observed and simulated AGN populations have been relatively limited. Here, we have used the cosmo-OWLS suite of cosmological hydrodynamical simulations to produce realistic synthetic catalogs of X-ray AGN out to $z=3$, with the aim of comparing the catalogs to the observed X-ray AGN population in the XXL survey and other recent surveys.  We focused on the unabsorbed X-ray luminosity function (XLF), the Eddington ratio distribution, the black hole mass function, and the projected clustering of X-ray AGN.
To compute the unabsorbed XLF of the simulated AGN, we used recent empirically-determined (luminosity-dependent) bolometric corrections, in order to convert the simulated bolometric luminosity into an observable X-ray luminosity. We show that, using these corrections, the simulated AGN sample accurately reproduces the observed XLF over 3 orders of magnitude in X-ray luminosity in all redshift bins from $z=0$ out to $z=3$.
To compare to the observed Eddington ratio distribution and the clustering of AGN, we produced detailed `{\it XMM-Newton}-detected' catalogs of the simulated AGN. This requires the production of synthetic X-ray images extracted from light cones of the simulations, which self-consistently contain both the X-ray AGN and the emission from diffuse, hot gas within galaxies, galaxy groups, and clusters and that fold in the relevant instrumental effects of {\it XMM-Newton}.  We apply a luminosity- and redshift-dependent obscuration function for the AGN and employ the same AGN detection algorithm as used for the real XXL survey.  We demonstrate that the detected population of simulated AGN reproduces the observed Eddington ratio distribution and projected clustering from XXL quite well. Based on these comparisons, we conclude that the simulations have a broadly realistic population of AGN and that our synthetic X-ray AGN catalogs should be useful for interpreting additional trends (e.g. environmental 
dependencies) and as a helpful tool for quantifying AGN contamination in galaxy group and cluster X-ray surveys.}


\authorrunning{E. Koulouridis et al.}
\titlerunning{The XXL survey - XIX}

\maketitle

\section{Introduction}

The cosmological evolution of supermassive black holes (SMBH) is a
vibrant topic in modern astrophysics. 
Its importance has been recognized ever since the discovery that virtually all massive galaxies in the local Universe host a central SMBH with a
mass proportional to that of the galaxy spheroid \citep[e.g.][]{Kormendy95,Magorrian98,Ferrarese00,Gebhardt00,Tremaine02,Marconi03,Gultekin09,Kormendy09,Zubovas12}. This tight relation indicates that SMBHs and
their host galaxies co-evolve, but the 
physical processes that lead to this relation are still debated. 

SMBHs grow primarily by accreting surrounding mass that leads to emission through various physical processes 
and to the appearance of an active galactic nuclei (AGN). An accurate
census of the AGN is essential in understanding the cosmic history of
accretion onto SMBHs and its relation to the host
galaxy. Theoretical models proposed an AGN-driven feedback
which can successfully expel gas from the galaxies in order to explain
this interactive co-evolution \citep[e.g.][]{Granato04,Monaco05,Springel05,Croton06,Hopkins06,Schawinski06,Cen11}. In addition, over the past decade or so
both semi-analytical models of galaxy formation and full cosmological hydrodynamical simulations faced the problem of
an excessively large number of bright galaxies formed in massive
haloes \citep[cooling crisis, e.g.][]{Balogh01}. These results
pointed towards the necessary inclusion of AGN feedback in order to
suppress the star formation and produce the observed luminosity
functions.

  AGN demographics can provide an assessment of the cosmic SMBH growth history.  The AGN luminosity function (LF) is an especially powerful tool when studied over a wide range of redshift and wavelength \citep[e.g.][]{Maccacaro83,Maccacaro84,Maccacaro91,Boyle93,Boyle94,Boyle00,Page96,Ueda03,Ueda14,Wolf03,Barger05, Hasinger05,LaFranca05,Richards06,Bongiorno07,Silverman08,Croom09,Aird10,Aird15,Buchner15,Assef11,Fiore12,Ranalli16,Fotopoulou16}.  Arguably, the most effective  way to detect active galaxies is through X-ray observations \citep[e.g.][]{Brandt15}. The majority of the detected extragalactic X-ray sources are AGN, while their unresolved integrated contribution essentially builds up the X-ray cosmic background \citep[][]{Setti89,Comastri95}.  Although several methods and models have been explored over the years, there are still uncertainties in the evolution of the LF at high redshift and the amount of nuclear obscuration. Further progress in such studies will require larger AGN samples and 
knowledge of the joint ($N_{\rm H},z$) distribution \citep[][]{Ueda14}.

Producing a realistic simulated X-ray AGN population that originates directly from the SMBH population can provide an invaluable tool in the study of structure and SMBH evolution. Used in conjunction with the underlying large-scale structure, it could hint to the physical mechanisms that lead to the observed properties of AGN populations, for example the correlation function of AGN, the halo occupation distribution (HOD), the environmental differences of obscured and unobscured AGN. It is also of great importance for X-ray cluster surveys, especially of the high-$z$ universe ($z>$1) where the level of contamination of the X-ray cluster emission by a powerful AGN is largely unknown.  Therefore, such a catalog can be of unprecedented value in the era of precision cosmology.  

However, difficulties arise from the many uncertainties regarding the observed X-ray properties of AGN.  Firstly, there is no established consensus on the ratio of X-ray to bolometric luminosity. Although several X-ray samples have been used over the years \citep[e.g.][]{Marconi04,Hopkins07,Vasudevan10,Lusso12,Shankar13} to produce a reliable bolometric correction function, the results remain discrepant. Secondly, the column density distribution of the AGN torus is a highly disputed topic. All X-ray background and unabsorbed X-ray luminosity function (XLF) studies \citep[e.g.][]{Ueda14,Buchner15,Ranalli16, Fotopoulou16} had to address the issue but the adopted approaches differ from study to study.  Most of the results do indicate, however, a strong luminosity dependence \citep[e.g.][]{Ueda03,Ueda14,Simpson05} and an evolution of the column density \citep[e.g.][]{LaFranca05,Hasinger05,Ueda14}.    

In the current paper, we have used the output AGN catalogs from the cosmo-OWLS suite of cosmological hydrodynamical simulations \citep[][]{Lebrun14} to produce a simulated population of X-ray AGN up to redshift 3, and we have compared with observations. As shown by \citet{Lebrun14}, models which include AGN feedback perform significantly better than those that do not with regards to reproducing the observed properties of local galaxy groups and clusters.  Assessing the realism of the predicted AGN population in the simulations is therefore a powerful independent test of the simulations that invoke AGN feedback.
In Sect. 2 we present the simulations and the SMBH modeling. We also present the XXL survey, which we used to compare the projected correlation function of the simulated AGN with observations, and within the framework of which the X-ray AGN modeling was undertaken. In Sect. 3 we describe the applied methodology and in Sect. 4 we present the results and compare the properties of the simulated AGN catalog with observations. Finally, in Sect. 5 we summarize our results and discuss the possible future applications of the catalog. 

 We call (a.) the X-ray AGN catalog after applying the bolometric corrections ``the unabsorbed X-ray AGN catalog", (b.) the one derived after applying obscuration ''the absorbed X-ray AGN catalog``, and (c.) the one after applying the simulated {\it XMM-Newton} observational features and the detection pipeline ``the detected X-ray AGN catalog". An outline of the procedure and of the products is presented in Table 1. When referring to the soft and the hard band we always mean the 0.5-2 keV and the 2-10 keV bands, respectively.

\section{Data description}

\subsection{The cosmological hydrodynamical simulations}

The cosmo-OWLS simulations were carried out with a version of the Lagrangian
TreePM-SPH code GADGET3 \citep[last described in][]{Springel05}, which
has been modified to include additional sub-grid physics. The volume was defined by a 400 $h^{-1}$ (comoving) Mpc on a side periodic box.
The initial conditions were based either on the maximum-likelihood cosmological parameters derived from the 7-year WMAP \citep[][]{Komatsu11} or the Planck data \citep[][]{PlanckXVI}. The number of particles was 2$\times1024^3$, yielding dark matter and (initial) baryon particle masses of $\sim 3.75\times10^9 h^{-1} M_{\sun}$ and $\sim 7.54\times10^8 h^{-1} M_{\sun}$ for the WMAP7 cosmology. In the current work, we have used the WMAP7 runs by default, since the WMAP-predicted cluster density is consistent with the observed number count in the XXL survey, in contrast to the Planck cosmology predictions \citep{Pacaud16}. 
Nevertheless, the Planck runs were also tested and a comparison is presented. Further details about the way radiative cooling rates, reionization, star formation, stellar evolution and SN feedback were implemented in the cosmo-OWLS can be found in \citet[][]{Schaye10} and references therein.

For each simulation, ten different light cones were produced, each of 25 deg$^2$, thus matching the area of one XXL survey field (see Sect. 2.3). The interested reader can refer to \citet{McCarthy14} for further details of the light-cone making method. X-ray maps for the hot diffuse gas were produced for each light cone by summing the X-ray emission of each gas particle along the line of sight in pixels of $2.5''$, matching the real XXL pixel scale. A description of how the X-ray emission of gas particles was computed can be found in \citet[][]{Lebrun14}. X-ray AGN were then added to the maps, using the actual locations of accreting SMBHs in the simulations (i.e. we create light cones for the SMBHs as well) and their predicted X-ray emission, which is described below.

\subsection{SMBH modeling in the cosmo-OWLS}

Three of the cosmo-OWLS runs included AGN feedback as the result of accretion onto SMBHs. This was incorporated
using the sub-grid prescription of \citet[][]{Booth09}, where the interested reader can find all the details of the modeling.  However, we summarize the essential ingredients for the present study.

During each simulation, an on-the-fly friends-of-friends (FoF) algorithm is applied on the dark matter particles. All haloes
with more than 100 particles (a corresponding mass of log$_{10}[M_{FoF} (M_{\sun}h^{-1})]\approx11.6)$ are seeded with SMBH sink particles.
The initial SMBH mass is 0.001 times the (initial) gas particle mass ($\sim10^5 M_{\sun}h^{-1})$. 

The simulated SMBHs grow via Eddington-limited, modified Bondi-Hoyle-Lyttleton accretion \citep[][]{Bondi44,Hoyle39} and by merging with other SMBHs. The accretion rate is given by:

\begin{equation}
\dot m_{acc}=\alpha\frac{4\pi G^2M^2_{\rm SMBH}\rho}{(c^2_s+u^2)^{3/2}},
\end{equation}
where $M_{\rm SMBH}$ is the mass of the black hole, $\rho$ and $c_s$ are the gas density and the sound speed of the local medium, and $u$ is the relative velocity of the black hole to the ambient medium.  The relation is modified with respect to the standard Bondi accretion rate through the inclusion of the multiplicative $\alpha$, that was originally introduced by \citet[][]{Springel05b} to correct for the limitations of the simulations.  Specifically, in typical cosmological hydro simulations, the numerical resolution is too low to resolve the Bondi radius, and therefore the estimated accretion rate will be an underestimate of the true rate.  Furthermore, and more importantly, many cosmological hydro simulations (such as OWLS, Illustris, EAGLE, etc.) do not include an explicit modeling of the cold interstellar medium (ISM), but instead invoke an equation of state for dense gas, in order to avoid numerical fragmentation.  The use of an equation of state, which adds pressure to the gas (to mimic 
turbulence in the ISM), can also lead to a significant underestimate of the gas density near the SMBH, and therefore an underestimate of the accretion rate onto the SMBH.

In order to overcome these problems, \citet[][]{Springel05b}, and most subsequent studies that used this model, adopted a constant $\alpha$=100. OWLS and cosmo-OWLS adopted a somewhat different strategy, following \citet[][]{Booth09}.  In particular, in \citet{Booth09}, $\alpha$ depends on the local gas density, as $\alpha \propto \rho^2$.  However, at low densities, which can be resolved by the simulations, the accretion rate reverts back to the standard Bondi rate (i.e. with $\alpha=1$).

The black hole mass grows following the relation:
\begin{equation}
\dot M_{\rm SMBH}=m_{acc}(1-\epsilon_r),
\end{equation}
where $\epsilon_r$ is the radiative efficiency of the black hole, fixed at 10\% here.  In addition, 15\% of the radiated energy is coupled to the surrounding medium (i.e. feedback), while the remaining 85\% is allowed to escape. 

The accretion rate is always limited by the Eddington rate:
\begin{equation}
\dot m_{\rm Edd}=\frac{4\pi G^2M_{\rm SMBH}m_p}{\epsilon_r\sigma_{T} c},
\end{equation}
where $m_p$ is the proton mass, $\sigma_T$ is the Thomson cross-section and c the speed of light. 

The Eddington ratio $\lambda$ is defined as
\begin{equation}
\lambda=L_{bol}/L_{Edd},
\end{equation}
where $L_{Edd}=(M_{\rm SMBH}/M_{\sun})\times1.3\times10^{38}$erg sec$^{-1}$.

Finally, SMBH-SMBH mergers takes place when two black holes were within a distance $h_{BH}$ and their relative velocity $\upsilon$ was less than the circular velocity ($\upsilon<\sqrt{Gm_{BH}/h_{BH}}$, where $h_{BH}$ is the smoothing length and $m_{BH}$ is the mass of the most massive SMBH). When these conditions are met, the merger takes place instantaneously.

%However, in this case, the modeling is limited to these two criteria and the process is instantaneous, since the simulations lack the resolution for any further analysis.    

\subsubsection{AGN feedback}

As discussed earlier, AGN feedback is an important ingredient of the simulations which is necessary to suppress star formation and avoid the excessive formation of very massive galaxies. \citet[][]{Lebrun14} showed that the inclusion of AGN feedback leads to good agreement between the stellar masses of real and simulated brightest cluster galaxies (BCGs). The feedback also regulates the accretion onto the black holes themselves.  Therefore, we anticipate that different feedback models will directly affect the predicted AGN demographics (e.g. the XLF). We note that SN feedback is also modeled in the simulations. In this section we summarize briefly the AGN feedback modeling.

cosmo-OWLS transforms a fraction of the rest-mass energy of the accreted gas into heating of the neighbouring gas particles, by increasing their
temperature. An advantage of the \citet[][]{Booth09} model is that it overcomes the problem of numerical overcooling (i.e. the problem that feedback energy can be rapidly radiated away due purely to low mass resolution).  This is accomplished by raising the temperature of only a small number $n$ of surrounding gas particles by a predefined amount of $\Delta T$. To this end, a fraction $\epsilon$ of the accreted energy is stored in the SMBH until it reaches the predefined value. $\Delta T$ and $n$ are chosen such as to produce a sufficiently long cooling time and the time needed for a feedback event to be shorter than the Salpeter time for Eddington-limited accretion.  It is shown that $\Delta T = 10^8$K and $n=1$ satisfy the two constraints (AGN 8.0 model). However, in \citet[][]{Lebrun14} two more values of $\Delta T$ were tested, that is $3\times10^8$K (model 8.5) and $5\times10^8$K (model 8.7). The AGN 8.0 model proved more suitable for the purposes of that paper with Planck cosmology, while with WMAP7 the 
observational data tends to be bracketed by the AGN 8.0 and AGN 8.5 models. In the current work we have
tested both models.

Note than when $\Delta T$ is set to a higher value, more time is needed to accumulate the energy 
to heat the gas particle and we actually simulate more energetic bursts.  As already noted, the net efficiency $\epsilon$ is set to 0.015, which results in a good match to the normalization of the $z=0$ relations between SMBH mass and stellar mass and velocity dispersion, as well as to the observed cosmic SMBH density, as demonstrated by \citet[][]{Booth09} and \citet[][]{Lebrun14}. 

Finally, the cosmo-OWLS output SMBH catalog, which is the input SMBH catalog in the current study, provides the position, the redshift, the mass and the bolometric luminosity $L_{bol}$ for all SMBHs for the 25 deg$^2$ light cones up to redshift $z=$3.

\subsection{The XXL survey}

The XXL Survey is the largest {\it XMM-Newton} project approved to date ($>$6 Msec), 
surveying two $\sim$ 25 deg$^2$ fields with a median exposure of 10.4 ks 
and at a  depth of $\sim5\times10^{-15}$ erg sec$^{-1}$ cm$^{-2}$ in the [0.5-2] keV soft X-ray band
(completeness limit for the point-like sources). The two fields have extensive multi-wavelength coverage 
from X-ray to radio. A general description of the survey and its goals was published by \citet[][]{Pierre16}.
To date some 450 new galaxy clusters have been detected out to redshift $z\sim2$, as well as more than 20000 AGN 
out to $z\sim4$. The main goal of the project is to constrain the dark energy equation 
of state parameter, $w$, using clusters of galaxies. This survey will also have lasting legacy value for cluster scaling laws 
and studies of galaxy clusters, AGN, and X-ray background. The XXL-S (Southern) field, which we use in the current study, is one of two XXL fields, centered at RA=23$^{h}$30 and DEC=-55$^{d}$00.

\section{Methodology}

In the following sections we describe the procedure used to convert the output black hole catalog of the simulations to the final X-ray AGN catalog.

We preselected our sample so that only active black holes were included. To this end, we set an absolute accretion rate threshold of $10^{-6} M_\odot/$year \citep[][]{Ho08}, which corresponds to a bolometric luminosity cut of $\sim 5\times10^{39}$ erg s$^{-1}$. This cut eliminated almost one-third of the SMBH sample, but we note that SMBHs with luminosities below this threshold would not be detected with current surveys. Therefore, our cut was a conservative one.  We further assumed that all AGN with luminosities exceeding this threshold were X-ray emitters and therefore potentially detectable in X-ray surveys. This was a reasonable assumption because almost all identified AGN by optical, infrared, and radio techniques show X-ray AGN signatures \citep[see review on AGN demographics by][and references within]{Brandt15}. Therefore, X-ray emission seems to be almost universal, at least for the luminous AGN.  Nevertheless, it appears that a small number of intrinsically X-ray weak but luminous AGN 
does exist \citep[e.g.][]{Wu11,Luo14}. However, current studies indicate that they are so rare that their impact on demographic studies should be substantially small \citep[e.g.][]{Gibson08,Wu11,Luo14}.

An alternative strategy, which has been adopted in some previous theoretical studies \cite[e.g.][]{Rosas16}, would be to select which AGN will be X-ray emitters based on the predicted Eddington ratio.  The motivation for this comes from the fact that there is a known empirical correlation between the Eddington ratio and the predominant emission wavelength \citep[e.g.][]{Dai04,Saez08}.  Without an Eddington ratio cut, there is the potential that we will include low-Eddington rate sources (e.g. radio AGN) in our sample.  However, as we will show, recent observations suggest that X-ray AGN actually span a relatively wide range of Eddington ratios (which we will compare to; see Fig. 5 and Fig. 6), which means that there would be a strong possibility to exclude genuine X-ray emitters by adopting a fixed Eddington threshold (e.g. 0.01, as adopted in some previous studies).  This argues against adopting a fixed Eddington rate threshold.  Furthermore, we will show that, with our adopted luminosity cut, only a negligibly small 
fraction of our selected simulated AGN have very low Eddington accretion rates of $\lambda < 10^{-4}$, which are typical of radio AGN.  

Below we describe the (inverse) bolometric corrections (i.e. to convert the simulated bolometric luminosity into an observable X-ray luminosity) and the application of AGN obscuration to produce our final X-ray AGN sample.

\begin{table}
\begin{minipage}{87mm}
\centering
\caption{Methodology outline}
\tabcolsep 3 pt
\renewcommand{\arraystretch}{1.8}
\begin{tabular}{|l|c|c|}
\hline
Tool or methodology  &  output  & results\vspace{-7pt} \\
{\em (1)}&{\em (2)}&{\em (3)}\\
\hline 
cosmo-OWLS (Sect. 2) & SMBH catalog    &  \\  
\hline
\multirow{2}{85pt}{bolometric corrections (Sect. 3.1)} &  \multirow{2}{65pt}{\centering unabsorbed X-ray AGN catalog}& \multirow{2}{65pt}{\centering unabsorbed X-ray LF (Sect. 4.1)} \\
 &    & \\ 
\hline
\multirow{3}{85pt}{absorption function (Sect. 3.2)}&  \multirow{3}{65pt}{\centering absorbed X-ray AGN catalog} &  \multirow{3}{70pt}{\centering Eddington ratio distribution \& black hole mass function (Sect. 4.2)}\\       
&&\vspace{-2pt}\\
&&\\
\hline 
\multirow{3}{85pt}{{\it XMM-Newton} instrumental effects (Sect. 3.3)}&  \multirow{3}{65pt}{\centering detected X-ray AGN catalog} &  \multirow{3}{77pt}{\centering projected correlation function (Sect. 4.3)}\\       
&&\vspace{-7pt}\\
&&\\
\hline
\end{tabular}
\tablefoot{{\em (1)} The applied tool or methodology (the sections where they are described) {\em (2)}, name of the output catalog,
{\em (3)} the results (the sections where they are described) } 
\end{minipage}
\end{table}

\subsection{Bolometric correction}

Despite a concerted effort to combine various X-ray and optical surveys (e.g. XMM-COSMOS, CDF-N, CDF-S, ROSAT, SDSS, 2dF) while exploiting the area of shallow surveys and the depth of pencil-beam surveys, there is still no general consensus between different studies 
on the fraction of the total bolometric luminosity ${\rm L_{bol}}$ that is emitted at X-ray wavelengths \citep[for a comparison between different studies see][L12 hereafter]{Lusso12}. 
Nevertheless, most studies do agree that the correction depends on the luminosity itself, in the sense that the correction becomes increasingly large with increasing bolometric luminosity. However, the scatter in published relations is relatively large. In addition, a number of studies \citep[e.g.][]{Vasudevan07,Vasudevan09b,Vasudevan10} presented evidence that the bolometric corrections depend primarily on the Eddington ratio and not the luminosity of their low-$z$ AGN samples. \citet[][]{Shankar13} studied thoroughly this relation using semi-empirical models of AGN, but they concluded that their modeling, although it becomes very elaborate, cannot reproduce well the observational constraints. We note, however, that L12 reported a clear correlation of increasing Eddington ratio with increasing luminosity up to redshift 2.3, which implies that probably both are correlated with the bolometric corrections in a similar way.

In the current study we have implemented the simple approach of adopting luminosity-dependent bolometric corrections only, of which we tried several.  As we will show, for recently-determined bolometric corrections from either L12 or \citet[][M04 hereafter]{Marconi04}, the simulations predict a hard XLF that is consistent with observations;  \citet[][]{Ranalli16}, \citet[][]{Aird15}, \citet{Miyaji15} and \citet{Buchner15} (see Sect.  4.1).

It is worth noting that we also explored using the bolometric corrections proposed by \citet[][]{Hopkins07}, but found significantly worse agreement with the observed XLF.  To estimate the bolometric corrections, \citet[][]{Hopkins07} combined a large number of optical, soft and hard X-ray, and mid-IR catalogs. They provide the bolometric corrections for a wide range of bolometric luminosities. However, we found that the level of the proposed corrections is very high, producing an under-luminous simulated X-ray AGN population that fails to reproduce the hard band unabsorbed XLF.  This may be attributed to the inclusion of reprocessed emission in their calculations (although we cannot rule out that the discrepancy could also be due in part to inadequacies in the underlying predicted bolometric LF).  M04, by contrast, constructed a template spectrum to study the local black hole properties of optical QSOs and they explicitly removed the IR bump in order to estimate the bolometric corrections without the 
reprocessed radiation.  However, they assumed that the template spectrum, and thus the derived bolometric corrections, is redshift independent. On the other hand, L12 derived empirical bolometric corrections using {\it XMM-COSMOS} hard X-ray selected AGN. Their corrections are generally smaller than those proposed by M04, but consistent within the scatter. The sample used in L12 spans the full redshift range up to $z=3$ but, as expected, the AGN population at low redshifts is undersampled. Therefore, it is possible that there is a mild evolution of the bolometric corrections which can reconcile the differences in the corrections proposed by L12 and M04.  In any case, we explore using both corrections in Sect. 4.1, showing that adopting either leads to reasonable agreement with the observed XLF. In both cases the functions are approximated by third degree polynomials:
\begin{equation}
y=\alpha_1x+\alpha_2x^2+\alpha_3x^3+\beta,
\end{equation}
where $y={\rm log_{10}}[L_{bol}/L_{band}]$, and $x={\rm log_{10}}[L_{bol}/L_{\sun}]-12$. The set of parameters ($\alpha_1, \alpha_2, \alpha_3, \beta $) are given by (L12: 0.217, 0.009, -0.010, 1.399) and (M04: 0.22, 0.012, -0.0015, 1.65) for $L_{band}=L_{[0.5-2 keV]}$, and by (L12: 0.230, 0.050, 0.001, 1.256) and (M04: 0.24, 0.012, -0.0015, 1.54) for $L_{band}=L_{[2-10 keV]}$.  

\subsection{Obscuration}

Obscuration was implemented for our X-ray catalog following the absorption function $f(L_X,z;N_{\rm H})$ introduced by \citet[][]{Ueda14}. To derive this function they used a highly-complete sample compiled from several surveys using {\it Swift/BAT, MAXI, ASCA, XMM-Newton, Chandra}, and {\it ROSAT}.  The function takes also Compton-thick AGN (log $N_{\rm H}>24$) into account. The level of absorption is strongly luminosity-dependent and it evolves with redshift. In particular, the frequency of absorbed AGN (log$N_{\rm H}>22$) rises steeply with decreasing AGN luminosity, rising from $\sim$20\% for high-luminosity AGN ($L_X>10^{45}$ erg sec$^{-1}$) to more than 80\% for the low-luminosity sources. Also, the function includes a positive evolution of the absorbed fraction with redshift, as reported by several studies \citep[e.g.][]{LaFranca05,Ballantyne06,Treister06,Hasinger08}. We note that there are large uncertainties involved in these calculations, as clearly stated in \citet[][]{Ueda14}, 
especially for the faint AGN. 

\begin{figure}[h!t]
\centering
\resizebox{7.5cm}{20cm}{\includegraphics[angle=0, origin=c]{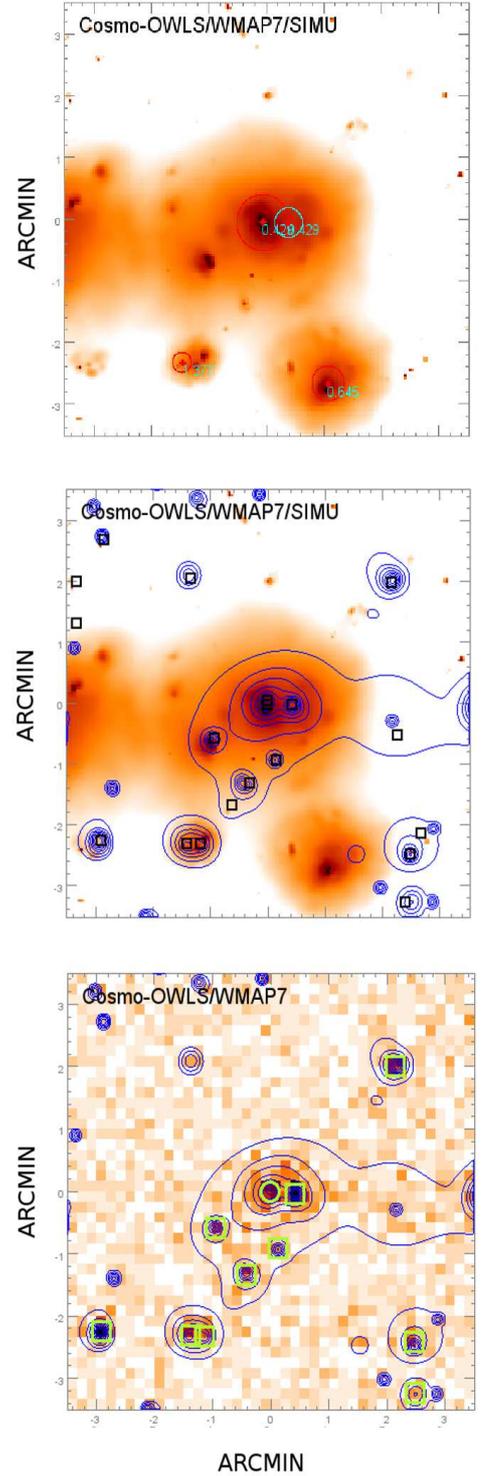}}
\caption{Simulated {\it XMM-Newton} images and source detection. Top: X-ray photon map from cosmo-OWLS overplotted with red (cyan) circles that mark the position and redshift of the input dark matter haloes (secondary haloes). The radius of each circle represents the $r_{500}$ radius. Middle: same as top overplotted with the X-ray contours (10 ks exposure) and the position of the input simulated AGN (black squares). Bottom: same as top after including {\it XMM-Newton} instrumental effects and background. Green squares (circles) mark significant detections of point-like (extended) sources by the detection algorithm.}
\end{figure}

\begin{figure*}[t]
\centering
\resizebox{19cm}{15cm}{\includegraphics[angle=270, origin=c]{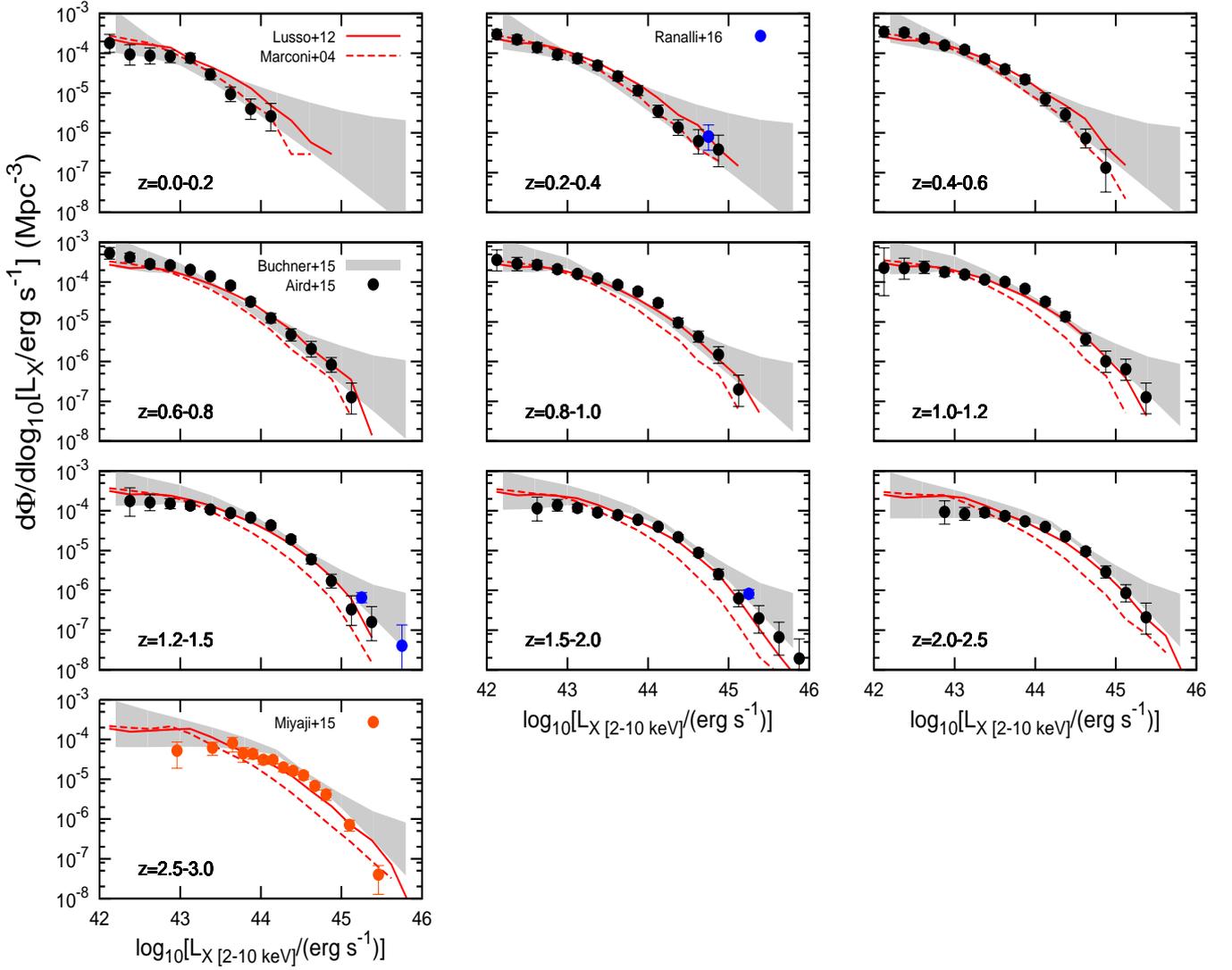}}
\caption{2-10 keV unabsorbed X-ray luminosity functions of synthetic and observed AGN. The eleven panels correspond to redshift bins up to redshift 3. The results of our modeling with cosmo-OWLS data are marked with red lines (continuous for bolometric corrections based on L12, dashed for M04). Black circles (with 1-$\sigma$ errors) denote the intrinsic hard XLF by \citet[][]{Aird15}. In the last redshift bin we plot data from \citet[][orange circles]{Miyaji15}, which span a more pertinent redshift range. The grey bands indicate the 90\% confidence interval of a non-parametric fit of observational data by \citet[][]{Buchner15}. For comparison, we also plot data points (blue circles) by \citet[][]{Ranalli16} (with 1-$\sigma$ errors) at the bright-end of the XLF (the 11 deg$^2$ of the XMM-LSS survey were used).}
\end{figure*}

\begin{figure*}[t]
\centering
\resizebox{19cm}{15cm}{\includegraphics[angle=270, origin=c]{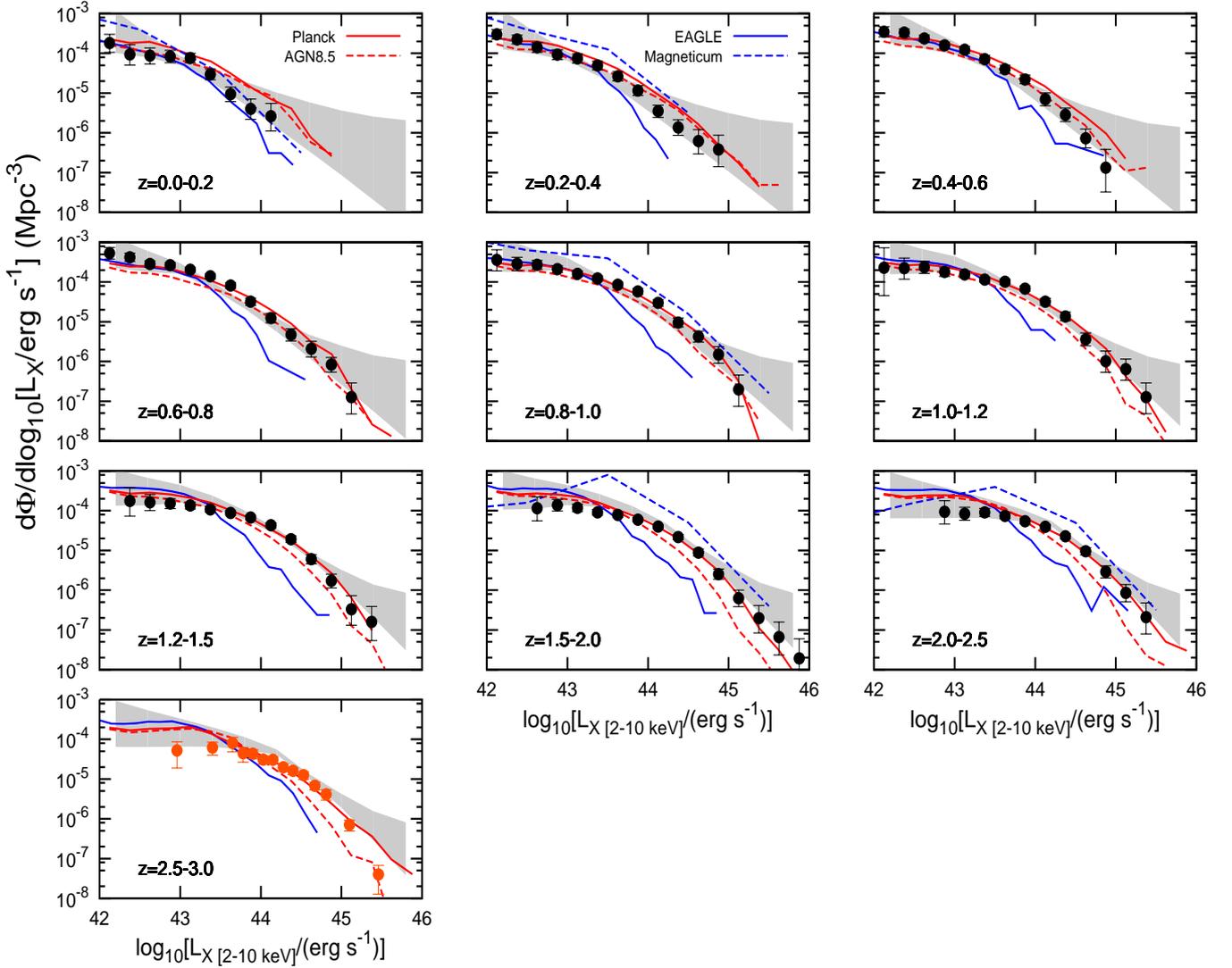}}
\caption{2-10 keV unabsorbed luminosity functions obtained using various cosmologies and AGN feedback models (see Sect. 2.2.1). We also overplot the results of similar analyses with the EAGLE (blue continuous lines) and the Magneticum Pathfinder simulations (blue dashed lines). Planck cosmology results (red continuous line) use the AGN8.0 model. Observed XLFs are plotted as in Fig. 2.}
\end{figure*}

\begin{figure*}[t]
\centering
\resizebox{19cm}{7cm}{\includegraphics[angle=270, origin=c]{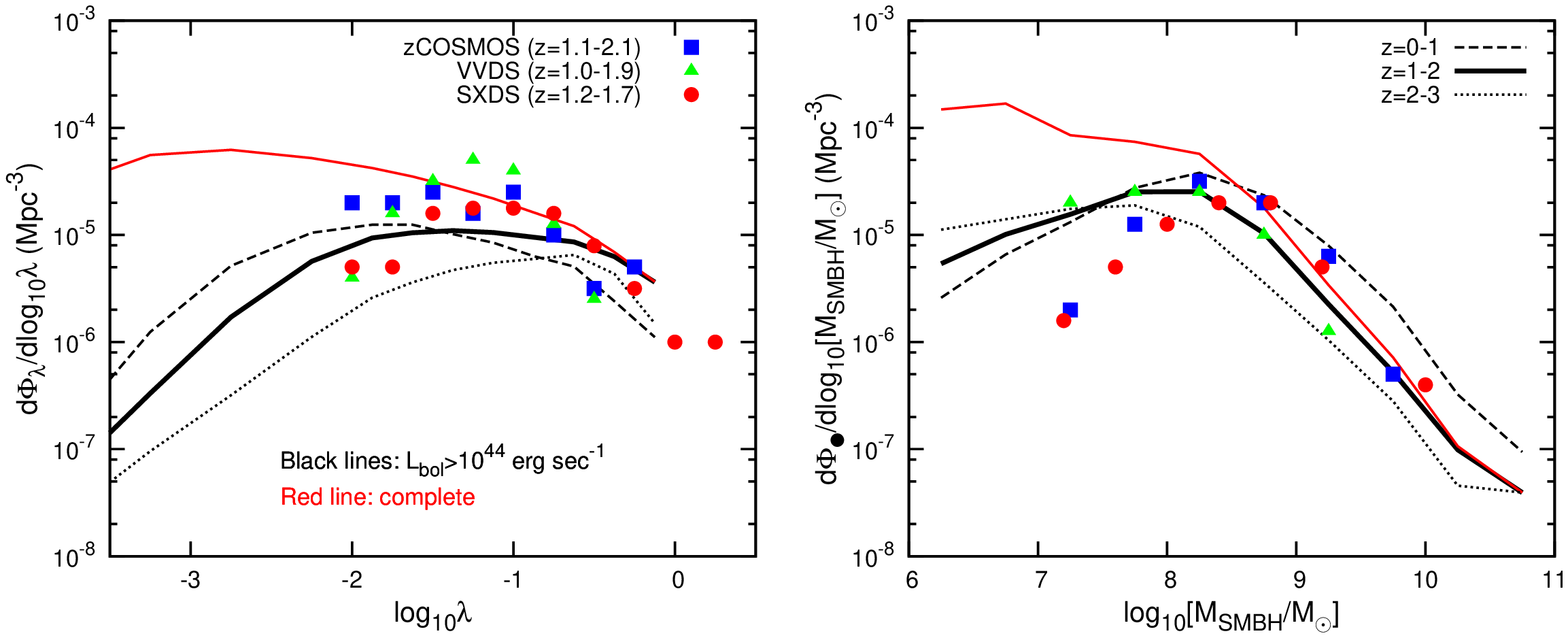}}
\caption{ERDF (left panel) and BHMF (right panel) of broad-line AGN (type 1, $N_{\rm H}<10^{22}$ cm$^{-2}$) between redshift 1 and 2 (black continuous line). We overplot relevant X-ray selected data from SXDS (red circles) and optically selected data from VVDS (green triangles) and zCOSMOS (blue squares), corrected for incompleteness with the 1/$V_{\rm max}$ method. A luminosity limit of $L_{bol}>10^{44}$erg sec$^{-1}$ was imposed on the simulation data according to the respective limitations of the above surveys. When the limit is relaxed (red line), the number of sources continuously increases towards low $\lambda$ and low $M_{\rm SMBH}$. We also present the respective distributions of sources with $z<1$ (dashed line) and $z>2$ (dotted line).}
\end{figure*}

We did not implement any further criteria that may play a role in the obscuration of black holes; for example interactions and merging of the host galaxies.  This could in principle have an impact on the correlation function of obscured AGN compared to the unabsorbed population.  However, studies using X-ray selected samples \citep[e.g.][]{Coil09,Ebrero09,Mountrichas12} did not find significant differences, although \citet[][]{Elyiv12} reported different clustering for hard and soft X-ray sources.

Obscured fluxes in the soft and the hard X-ray bands were calculated with NASA's HEASARC tool PIMMS\footnote{https://heasarc.gsfc.nasa.gov/docs/software/tools/pimms.html} (Portable, Interactive Multi-Mission Simulator), where the k-correction was applied assuming a power law spectra with photon index $\Gamma=1.9$ \citep[e.g.][]{Nandra94}.

\subsection{Simulated {\it XMM-Newton} images and source detection}

Synthetic X-ray images were created from the perfect-sky X-ray photon-maps and
the input X-ray AGN catalog. We also added a realistic background, which
included X-ray photons (vignetted), solar soft protons (vignetted),
and particles (not vignetted). We modeled the photon background, following \citet[][]{Snowden08},
as the sum of a Galactic and an extragalactic contribution. 
The Galactic contribution was computed by the superposition
of two absorbed MEKAL components \citep[][]{Mewe85} at $0.1$ keV and $0.25$ keV
from the galactic halo and another, unabsorbed, MEKAL component at $0.1$ keV
from the Local Hot Bubble; the extragalactic contribution (from unresolved AGN)
was modeled as a power law with index 1.46. The solar soft proton background was modeled after \citet[][]{Snowden08}, as a power law
with index 0.9; particle background was computed from 200ks {\it XMM-Newton} exposures with
closed filter wheel and we chose not to include flares.

Finally, an ideal event list was created by merging the above contributing photons. It was then blurred
to simulate the {\it XMM-Newton} instrumental effects: PSF blurring (assuming a King profile PSF), energy blurring,
vignetting; particle background was also added.
In all cases we assumed a 10 ks exposure time, as in the XXL survey. Photons were reshuffled in position and energy, or were discarded according to the simulated
local effective area, exposure time, vignetting factor, detector (MOS1, MOS2, PN) or filter (THIN). 
Therefore, we obtained three event lists (one for each EPIC detector) that included instrumental effects and
which were converted to images in the 0.5-2 keV and 2-10 keV bands at $2.5''$ per pixel. We also produced the corresponding exposure maps.

Source extraction was performed on these images for the soft
and the hard band separately, in the same way as for the XXL survey images, via the
XAmin pipeline \citep[][]{Pacaud06}. In more detail,
first a preliminary list of source candidates was selected by running
SEXtractor \citep[][]{Bertin96} on a wavelet smoothed combined (MOS1, MOS2, PN)
X-ray image. Then, on each candidate source, a series of fits was performed on the three raw X-ray images:
a point source model (assuming a position-dependent {\it XMM-Newton} PSF), an extended source model (assuming a  $\beta=2/3$ profile),
a double point source model (two {\it XMM-Newton} PSFs close on the image), and an extended$+$point source model
($\beta=2/3$ profile with central {\it XMM-Newton} PSF). In \citet[][]{Pacaud06} the threshold level for a significant detection has been chosen in order that any detection compatible with a non-extended source would have a $\sim$99\% probability of being a real source and not background fluctuation.

An example of the resulting images and pipeline detections of the above procedure is presented in Fig. 1. The detected AGN have usually more than 10 counts, while the remaining input sources are either detected at low significance or not detected at all.    

\section{Results}

In the following sections we present the comparison of the synthetic AGN catalogs with observational results. Obtaining a good agreement is essential for any further application of the simulated catalogs. 

\subsection{Unabsorbed hard X-ray luminosity function}

After implementing the bolometric corrections described in Sect. 3.1, we produced catalogs of X-ray AGN and their respective intrinsic X-ray luminosity (before obscuration). To assess how closely these catalogs relate to the observed X-ray AGN population, we compare our results to the unabsorbed (de-obscured) hard band XLF of \citet[][]{Ranalli16}, \citet[][]{Aird15}, \citet{Miyaji15} and \citet{Buchner15}. The differential luminosity function $\Phi$ is defined as the number of objects $N$ per comoving volume $V$ and per unabsorbed luminosity $L$ as follows:
\begin{equation}\label{eq:lf}
\Phi(L,z)=\frac{d^2N(L,z)}{dVdz}.
\end{equation}
The comparison within ten redshift bins up to $z=3$ is illustrated in Fig. 2. For clarity we mainly plot data points from \citet[][]{Aird15}, except in the $z=2.5-3$ range where \citet{Miyaji15} data are more pertinent. We also plot the 90\% confidence interval of the non-parametric fit by \citet{Buchner15}. This is an important addition since their analysis, which takes all uncertainties and the contribution of Compton-thick AGN into account, does not predict the sharp flattening of the XLF towards low-luminosity high-redshift bins, a common behaviour of previous parametric fits.    

Using the empirical bolometric corrections of L12, the simulations reproduce the observed XLF in all redshift bins, although there is possibly a slight overestimate for the local population at $z<0.5$ (Fig.2, top panels), according to the XLF by \citet[][]{Aird15}. However, the results are more consistent with \citet{Buchner15}.  Using the template spectra corrections of M04, the simulations reproduce the XLF up to roughly $z\sim0.5$, but somewhat underestimate it at higher redshifts.  Recall that the M04 corrections are probably more accurate for the low redshift population, since they were computed from a template spectra at $z$=0, while the L12 corrections are based on X-ray observations that cover the full redshift range but undersample the local population.  Therefore, assuming a mild evolution of the bolometric corrections, one can use the M04 functions for the low redshift sources ($z<0.5$) and L12 for high-$z$ sources.  Alternatively, L12 can be used exclusively, bearing in mind the probable 
overestimation of bright low-$z$ sources, although all points are consistent within 2-$\sigma$.  
We note that, applying a mild evolution on the M04 relation in order to reach the L12 level gradually by $z\sim0.5$ does not alter the results considerably.  We will therefore use the results based exclusively on the L12 estimations for the rest of the paper, although we thoroughly tested all alternatives. No qualitative differences were found.

In general, it is apparent that simulations are in good agreement with observations within all redshift and luminosity bins.
Nevertheless, above redshift 1.5 the simulated points in low-luminosity bins start to deviate, showing a tendency to overestimate the number of faint AGN. This discrepancy, which evolves with redshift, could be due to the limitations of the simulations, or the applied bolometric corrections, or the completeness of the observational surveys. However, we note that the simulations are fully consistent with the non-parametric results of \citet{Buchner15}, which do not support the sharp flattening of the XLF. For relatively shallow surveys like the XXL (10 ks average exposure time), this area of the XLF is mostly unprobed, since such faint sources at such high redshifts would not be detected.  However, it becomes more relevant for deeper surveys. At the bright end, our results agree very well with the XLF by \citet[][]{Aird15}, but they are located at the lower limit of the fit by \citet{Buchner15}. The plotted points of the XLF by \citet[][]{Ranalli16}, where they also use the 11 deg$^2$ of the XMM-LSS field, shows that we may indeed underestimate the bright population at high redshifts, but not greatly.

Finally, in Fig.3, we present the X-ray luminosity functions that we obtain using a different cosmology (Planck, as opposed to WMAP7) and the AGN8.5 feedback model from cosmo-OWLS (as opposed to our default choice, the AGN8.0 model).  We use L12 bolometric corrections.  It is apparent that changing the cosmology does not affect the results, since they are extremely similar to what we obtain with WMAP7 (Fig.2).  On the other hand, as expected, the AGN feedback plays an important role.  The relatively low level of the XLF for the AGN8.5 model, compared to the AGN8.0 model, shows that adopting a more powerful feedback results in a less effective accretion and therefore, in a less luminous AGN population.

We also compare our results with those of other recent simulations, including the EAGLE \citep{Rosas16} and the Magneticum Pathfinder simulations \citep{Hirschmann14} in Fig.~3.  In terms of the comparison to EAGLE, the predicted XLFs agree relatively well at the faint end of the XLF, while they tend to underpredict the bright end. This difference may be due to the limited volume of the EAGLE simulations, the use of the M04 bolometric corrections, the exclusion of low-$\lambda$ sources (they omit log$_{10}\lambda<-2$ sources), and/or differences in the modeling of SMBH accretion rates.  By contrast, the Magneticum Pathfinder simulations, which also use the M04 corrections, tend to overpredict the XLF at most luminosities and the discrepancy tends to grow with redshift.  The steep drop of the predicted XLF at high redshifts and low luminosities may be due to the adoption of an inefficient mode of accretion for all log$_{10}\lambda<-1$ sources.

\subsection{Eddington ratio and SMBH mass distribution}

\begin{figure*}[t]
\centering
\resizebox{18cm}{13cm}{\includegraphics[angle=270, origin=c]{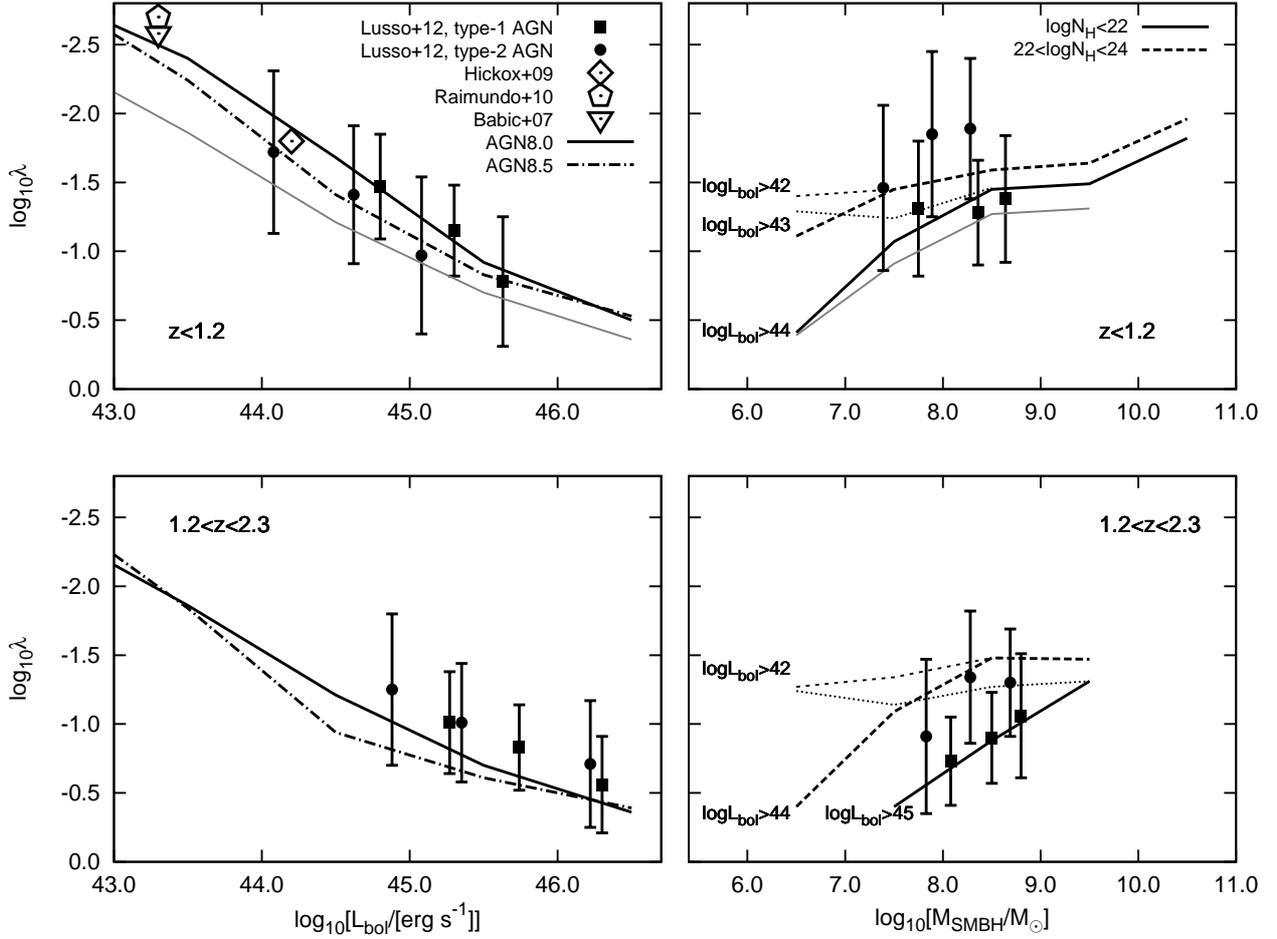}}
\caption{Eddington ratio vs. bolometric luminosity (left panels) and black hole mass (right panel). The simulated AGN sample was divided in two redshift bins, $z<1.2$ (top panels) and $1.2<z<2.3$ (bottom panels), and in unobscured (type-1, $N_{\rm H}<10^{22}$ cm$^{-2}$) and obscured (type-2, $10^{22}$ cm$^{-2}>N_{\rm H}>10^{22}$ cm$^{-2}$) sources, to match the L12 samples (circles and squares, mean values with 1-$\sigma$ errors). X-ray luminosity lower-limits (as marked on the plots) were also imposed for the same reason. When the lower-luminosity limits are relaxed the results are shown with dotted (type-1) and dashed lines (type-2). Obscuration does not affect the results plotted on the left, therefore only one line is drawn for the two samples. The gray lines are the respective results of the type-1 sample at the high-$z$ range, plotted to demonstrate the evolution. We also plot the results of the AGN8.5 model (dash-dotted lines) and the mean $\lambda$ values (not corrected for incompleteness) 
of previous studies at the mean luminosities of their samples. See Sect. 4.2 for more discussion on the observed trends.}
\end{figure*}

\begin{figure*}[t]
\centering
\resizebox{15cm}{15cm}{\includegraphics[angle=270, origin=c]{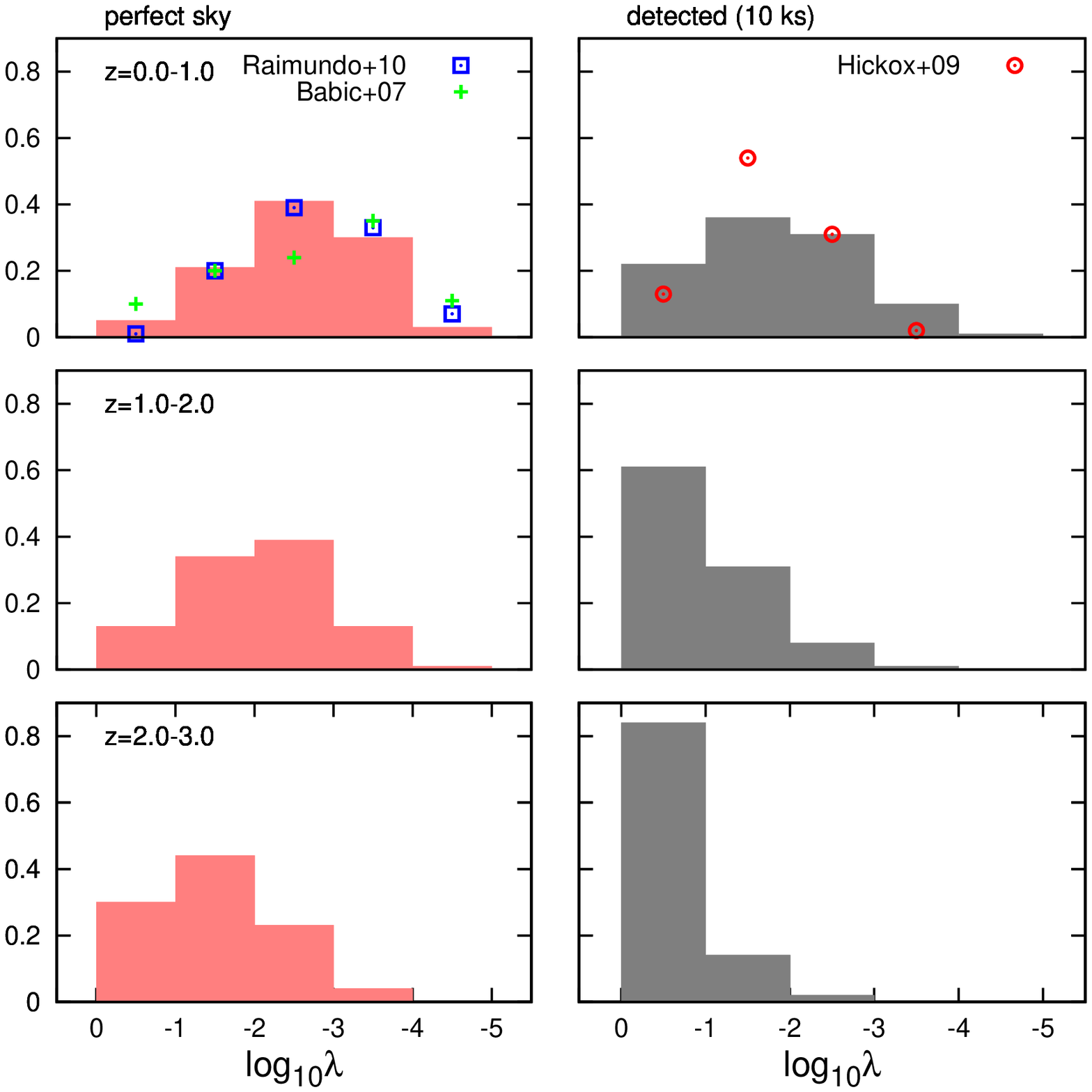}}
\caption{Eddington ratio distribution of the X-ray AGN within three redshift bins. On the left panels we plot the perfect-sky distribution and on the right panels the distribution after all observational and {\it XMM-Newton} instrumental effects were simulated (10 ks exposures, see Sect. 3.3). To illustrate observational selection effects, we overplot deep observational data on the perfect-sky distribution, and shallow on the 10 ks exposures (see Sect. 4.2 for more discussion).}
\end{figure*}

In this section, we study the differential Eddington ratio distribution function ($\Phi_\lambda$) and the differential SMBH mass function ($\Phi_\bullet$) of the synthetic X-ray population. The two functions follow the formalism of eq. (\ref{eq:lf}) replacing $L$ with $\lambda$ and $M_{\rm SMBH}$, respectively.

The Eddington ratio, being the ratio of the bolometric luminosity to the Eddington luminosity, is a clear indication of activity, although there is no explicit threshold which characterizes a turning point. Also, it is apparently redshift dependent. In the local Universe, the majority of AGN have $\lambda$ between $10^{-6}$ to $10^{-3}$ \citep[see review by][and references therein]{Alexander12}.  In the same review, they also argue that optically-detected AGN have an Eddington ratio distribution that peaks at $10^{-2}$.  On the other hand, X-ray AGN from z=0.3 to ~2.5 have a typical Eddington ratio between $10^{-4}$ to $10^{-1}$ \citep[e.g.][]{Babic07,Hickox09,Raimundo10,Lusso12}. However, at higher redshift the uncertainties are very large.

Firstly, we compare the intrinsic unobscured AGN population (type 1, $N_{\rm H}<10^{22}$ cm$^{-2}$), between redshift 1 and 2, to the results of relevant studies: X-ray selected sources ($z$=1.18-1.68) from the {\it SUBARU XMM-NEWTON Deep Field} \citep[SXDS,][]{Ueda08} described in \citet[]{Nobuta12}, and optically selected sources from the VVDS \citep{LeFevre13} and zCOSMOS \citep{Lilly07} surveys ($z$=1.0-1.9 and $z$=1.1-2.1, respectively) described in \citet{Schulze15}. To determine the unobscured simulated sample we apply the torus obscuration as described in Sect. 3.2. Comparing the unobscured population is the optimal choice, since obscuration corrections are minimal, especially in the hard X-ray band. In addition, a significant part of the accretion growth probably takes place within this redshift range. 
In Fig. 4, we plot the Eddington ratio distribution function (ERDF) and the black hole mass function (BHMF) of the above observational data and of our results (limited to $L_{bol}>10^{44}$ erg sec$^{-1}$). Observational data are corrected for incompleteness with the 1/$V_{max}$ method. We find a good agreement between simulations and observations in both cases. However, the shape of the VVDS ERDF is discrepant. We also plot the distribution of our data in the low-$z$ ($z<1$) and the high-$z$ ($z>2$) range. There is a clear evolution of the two functions, namely a significant increase of low-$\lambda$ and high-mass sources toward lower redshifts. Owing to the luminosity limit, the number of sources down to approximately $\lambda$=-2 increases only in the low-$z$ range and then rapidly decreases. However, if we relax the imposed luminosity limit, the number of sources increases continuously towards low $\lambda$ and low $M_{\rm SMBH}$, in agreement with the modeling of \citet[][]{Schulze15} which takes the low-flux sources below the limit of the surveys into account. 

Secondly, to reproduce the observational results presented in L12, we divide our sample in two redshift bins, $z<1.2$ and $1.2<z<2.3$, and in unobscured (type-1, $N_{\rm H}<10^{22}$ cm$^{-2}$) and obscured (type-2, $10^{22}$ cm$^{-2}>N_{\rm H}>10^{22}$ cm$^{-2}$) sources. X-ray luminosity lower-limits were also imposed for the same reason. In Fig.5 we plot Eddington ratio vs. bolometric luminosity (left panels), and black hole mass (right panels). There is an excellent agreement between simulations and observations within 1-$\sigma$. AGN8.5 results are more discrepant, especially in the high-$z$ range. We note that the axes are not independent and the trends need to be carefully 
explained. As expected, obscured and unobscured sources with the same intrinsic luminosities have the same Eddington ratio distributions (the same lines represent both samples in the left panels). The differences between the two types in the low-$z$ range, reported in L12, are not observed. Nevertheless, we find a clear evolution toward higher Eddington ratios at higher redshifts. On the other hand, when $\lambda$ is plotted versus mass the two AGN types differ. We argue that the difference is a result of the shift of the type-2 sample toward lower luminosities, meaning that if we select subsamples of the same luminosity distribution then the differences disappear. Nevertheless, the evolution is again apparent. Finally, if we relax the luminosity lower-limits, the simulated SMBH distributions flatten significantly, as expected by the shape of the BHMF in Fig. 4.    

In Fig. 6, we plot the Eddington ratio distribution of the simulated X-ray AGN catalog divided in three redshift bins. There is a clear increase of the of high-$\lambda$ fraction with increasing redshift, both before (left panels) and after (right panels) introducing the observational and instrumental effects described in Sect. 3.3 (10 ksec exposures). The low-$z$ AGN sample exhibits the lowest Eddington ratio values that peak roughly at $10^{-3}$ in both cases, while the majority of sources above $z$=2 have $\lambda$ values above 10$^{-2}$. Evidently, the steep evolution found for the detected sources is partly due to selection effects, since deeper surveys probe more low-$\lambda$ AGN at higher redshifts than shallow ones. This is demonstrated by overplotting data from the {\it Chandra} deep fields \citep[][]{Raimundo10,Babic07} and from the Galaxy Evolution Survey (AGES) \citep[][]{Hickox09}; the deep surveys trace the perfect-sky distribution, while the shallow match with our 10 ks exposures. This is in agreement with the strong positive correlation of $\lambda$ with luminosity, found in previous studies and presented in Fig.5.

Considering the above results, we conclude that our final AGN catalog follows the observed trends rather well.

\subsection{Projected correlation function and comparison with observations}

The final assessment of the simulated X-ray AGN catalog is the comparison of the predicted large-scale spatial distribution, as quantified by the projected two-point correlation function, with that of the real XXL data. This is of great importance since large-scale structure is a powerful diagnostic for
tracing the cosmic evolution of the AGN (and galaxy) populations.  We note that X-ray, IR and radio-selected AGN display different clustering properties, a fact which implies that specific modes of SMBH accretion may be related to the host dark matter halo \citep[e.g.][]{Hickox09,Melnyk13}, although selection effects cannot be ruled out.

The soft band projected correlation function of the southern XXL sample of
spectroscopically confirmed point-like sources
and its possible systematics will be presented in detail in a forthcoming paper.
The southern field has been chosen for this study due to the
homogeneity of its spectroscopic follow-up data, which is based uniquely on the
multifiber AAOmega facility on AAT, as compared to the
northern field which is based on a compilation of different surveys with
different instruments, limiting magnitudes, selection biases and solid angles. 
The XXL-S spectroscopic sample contains roughly $\sim 3740$ 
out of the $\sim4100$ total X-ray point
sources (a $\magcir$90\% completion) 
with $r$-band magnitude $\lesssim 21.8$ (the instrument
detection limit), obtained during two AAT observing runs.
The fraction of sources being stars is $\sim$10\%, and our final AGN
spectroscopic sample therefore consists of 3355 unique sources, out of
which 3106 are detected in the soft X-ray band sources and 1893 in the hard.

To compare the simulation with the XXL-S AGN projected 
correlation function, which is based only on confirmed sources, we 
need to avoid the spurious simulation detections of the pipeline. 
To this end, we
correlated the resulting catalog of significant pipeline detections with the
true simulated X-ray AGN input catalog (before the creation of the XMM
images). This resulted in $\sim7000$ soft band X-ray
sources, a number consistent with that of the real
XXL data but a factor of $\sim$2 larger than that of
the XXL-S sources with spectroscopy, an unavoidable fact due to the limiting
magnitude of the AAOmega spectroscopic facility.

To avoid the so-called redshift space distortion effects
 we used the 
projected correlation function, $w_p(r_p)$ \citep[][]{Davis83}, 
which is based on deconvolving the
redshift-based comoving distance, $s$, in a component parallel and
perpendicular to the line of sight, $\pi$ and $r_p$, respectively, as
$s^2=r_p^2+\pi^2$.
Then the so-called projected correlation function can be found
by integrating $\xi(r_p,\pi)$ along the $\pi$ direction:
\begin{equation}\label{eq:wp}
w_p(r_p)=2\int_{0}^{\infty}\xi(r_p,\pi) \mathrm{d}\pi \;.
\end{equation}
The real space correlation function can be recovered
according to Davis \& Peebles (1983):
\begin{equation}\label{eq:wp}
 w_p(r_p)=2\int_{0}^{\pi_{\rm max}}\xi\left(\sqrt{r_p^2+\pi^2}\right) 
{\rm d}\pi =2\int_{r_p}^{\infty}
 \frac{x \xi(x)\mathrm{d}x}{\sqrt{x^2-r_p^2}}\;.
 \end{equation}
Modelling $\xi(x)$ as a power law one obtains:
\begin{equation}\label{eq:wp_model}
w_p(r_p)=A(\gamma) r_p \left(\frac{x_{0}}{r_p}\right)^{\gamma},
\end{equation}
with $x_{0}$ the projected comoving clustering length at the effective 
redshift of the sample, and 
\begin{equation}
A(\gamma)=\Gamma\left(\frac{1}{2}\right)
\Gamma\left(\frac{\gamma-1}{2}\right)/\Gamma\left(\frac{\gamma}{2}\right),
\end{equation} 
with $\Gamma$ the usual gamma function.

We note that eq. (\ref{eq:wp}) holds strictly for $\pi_{\rm max}=\infty$,
while in order to avoid redshift-space distortions
the integral is performed up to a finite value of
$\pi_{\rm max}$, which in turn produces an underestimation of the underlying
projected correlation function. 
However, for the aim of comparing the clustering of the real XXL-S
sources to that of the simulated AGN we do not recover the true projected
comoving correlation length, $x_0$, but we just compare directly the $w_p(r_p)$
representation of the correlation function for the same value of
$\pi_{\rm max}$.

\begin{figure}[t]
\centering
\resizebox{8cm}{8cm}{\includegraphics{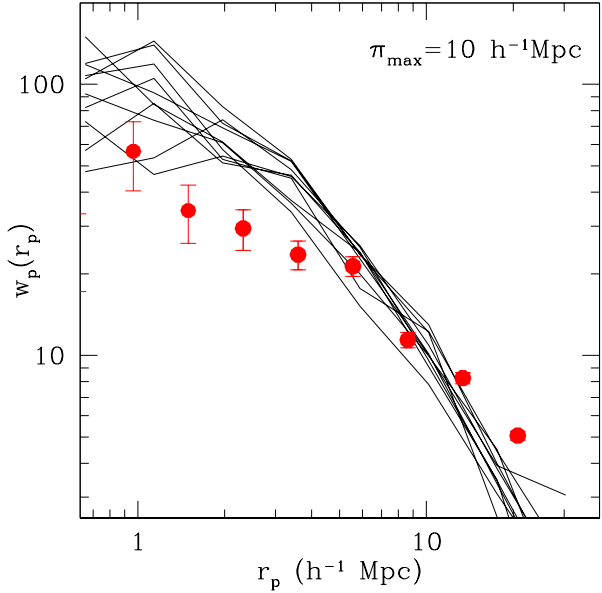}}
\caption{Projected correlation function of the XXL-S
  point-like sources (in red) and of the simulated X-ray AGN, detected
through the XXL pipeline (black lines, ten realizations).}
\end{figure}

In Fig. 7, we present the projected correlation function of the ten
realizations of the simulated XXL point-sources together
with that of the XXL-S spectroscopic sample. In both cases
we have limited the sources to those with $L_X>10^{41}$
erg sec$^{-1}$. It is evident that  there is a quite good consistency between
data and simulations for $r_p\gtrsim 3$ $h^{-1}$ Mpc, although at
small separations there is a deficiency of the XXL-S correlation function 
with respect to that of the simulations (a
fact which could possibly be attributed to the spectroscopic targeting
strategy which will be discussed in a forthcoming paper).

\section{Summary and discussion}

We presented the methodology used to produce a simulated population of
X-ray AGN from the SMBH population of the cosmo-OWLS
hydrodynamical simulations. The resulting AGN catalogs were compared
with observations to assess if they follow the
observed trends. We used ten light-cones of 25 deg$^2$ each, up to redshift 3.

Black holes in cosmo-OWLS grow through accretion of
surrounding gas and merging with other black holes.  Stellar
disruption is neglected.  Some previous studies argue, however, that it may play an
important role for AGN demographics \citep[e.g.][]{Milosavljevic06}, 
i.e. that many low-luminosity AGN may be due to the accretion of disrupted stellar mass. The rates of these events, as reported from X-ray surveys, are rather low ($10^{-4}-10^{-5}$/yr/galaxy) and agree well with theory and
simulations \citep[see][]{Komossa12}.
 Simulations however, showed that these rates are independent of the
 SMBH mass and thus only the growth of the intermediate or least massive
 SMBHs may be dominated by stellar disruptions \citep[e.g.][]{Brockamp11}, 
 while white dwarfs (extremely common) can only be observed
 in X-rays when they are disrupted by intermediate-mass black holes,
 $M_{\rm SMBH}<10^5 M_{\sun}$ \citep[e.g.][]{Luminet89,Rosswog09}. In addition, 
 observations show that the X-ray LF for
 moderate-luminosity active galactic nuclei is not due to tidal
 disruptions \citep[][]{Luo08}. We note that there are many
 uncertainties affecting these results and other effects which might
 reduce the fraction of stellar matter that is finally accreted by the
 black hole.  

 In the present study we have used the bolometric corrections calculated in
L12 from X-ray AGN in the XMM-COSMOS survey. M04 bolometric corrections, derived from template 
spectra, can also be used at the low-$z$ range. We argue that the two approaches are complementary (see Sect. 3.1).
Probably the most interesting result is how well the
simulated catalog reproduces the intrinsic luminosity
function \citep[][]{Aird15,Miyaji15} in almost all redshift bins and
luminosities.  A small discrepancy only appears at low-luminosities
above redshift 1.5, which increases with redshift. This discrepancy is
also present in other hydrodynamical simulations like the EAGLE
\citep[][]{Rosas16} and the Magneticum Pathfinder simulations
\citep[][]{Hirschmann14}.
However, we note that our results are in good agreement with the non-parametric XLF of \citet[][]{Buchner15}.

To produce obsured and unobscured AGN catalogs, we
applied obscuration to all our sources following the obscuration function 
by \citet[][]{Ueda14}. Following the observational trends, the function is luminosity-dependent and it evolves with redshift. 
Additional induced obscuration during galaxy merging was
not considered.  However, it is possible that a correlation of AGN obscuration with merging
exists, meaning that galaxy interactions and merging may lead to the triggering of SMBH activity 
\citep[e.g.][]{Hopkins08,Koulouridis06b,Koulouridis06,Koulouridis13,Villarroel14},
and to an enhancement of obscuration during the initial stage of AGN 
evolution \citep[e.g.][]{Koulouridis14,Villarroel17}.
  
We compared our AGN catalog properties with observational results (Eddington ratio distribution, black hole mass function) and we concluded that the  
simulated AGN population comprises sources that reproduce well the observed tendencies and the
evolution of the Eddington ratio, meaning that at higher redshift AGN accrete
more efficiently. Selection effects were also discussed.

We also compared the projected two-point correlation function of the
simulated AGN catalog with the corresponding one from the $\sim$25 deg$^2$
southern XXL field. 
The relatively good reproduction of the X-ray AGN large-scale
structure, both in observations and the simulation, has important consequences for
cosmology as it is related to the initial fluctuation spectrum and its
evolution. It further implies that the dark
matter haloes, hosting X-ray selected AGN, correspond directly to the
simulated ones, and thus the simulation provides a test-bed for
understanding the physical processes shaping the triggering and
evolution of the SMBHs in the Universe. We caution that the selection of the sources is not exactly the same, 
with the XXL-S data sources being a magnitude limited sample defined by the AAOmega limit of $r\simeq 21.8$.
Nevertheless, another interesting part of the general agreement is the fact that an
optical host-galaxy magnitude limited AGN sample agrees quite well with the underline X-ray AGN sample,
represented by the simulation data. In a forthcoming paper (Plionis et al. in prep.), 
which studies the AGN clustering in much greater detail, 
we perform a thorough and consistent comparison of the simulations and the XXL point-source redshift data.

On the X-ray cluster side, this sample can give valuable insight
for the high redshift ($z>1$) X-ray cluster population. X-ray clusters
are indeed detected in the redshift range between $z$=1 and 2, but the
level of AGN contamination and their selection function are completely
unknown. Very little is also known for the AGN which reside in
clusters (not the BCG) at these redshifts. There are indications of a
turn-over point at $z=1$ where not only AGN \citep[e.g.][]{Martini13}
but also star forming galaxies behave differently regarding their preference on
dense environments. Our catalogs are well suited to explore this
kind of questions in a statistical sense.

On the other hand, a successful synthetic AGN population should reproduce not only the observed AGN demographics, but also the detailed scaling relations of SMBHs, including their slope, amplitude, intrinsic scatter, and evolution. Recent studies demonstrated the essential role of the velocity dispersion in the relation between SMBHs and their host galaxies \citep[e.g.][]{Bluck16,Shankar16}. In addition, there is evidence of significant bias in the Maggorian relation \citep[e.g.][]{Lasker16,Reines16,Shankar16}, which introduces further complications for a realistic AGN modeling. Unfortunately, the relatively low resolution of the current simulations (a spatial resolution of 4 kpc/h, which owes to the fact that we are simulating huge volumes of the universe in order to model the galaxy cluster population) prevents us from being able to make meaningful comparisons of this sort at present. Measurements of the line-of-sight velocity dispersions at small scales would therefore be unreliable. 
Furthermore, we note that these simulations, like most cosmological simulations, do not reproduce in detail the observed galaxy stellar mass function, therefore we do not expect some of the scaling relations to be realistic.

In the present study, we have focused on the quasar demographics first, as this is crucial to our modeling and interpretation of the XXL survey. Going forward, however, the models must continue to be improved and challenged.

Given the limitations of the simulations and the uncertainties of the models 
used in the current work, we were able to produce synthetic X-ray AGN catalogs
which perform well when compared with observations.
The advantage of these catalogs is that the properties of the X-ray sources 
are directly linked to that of their host dark matter haloes and thus 
they can be used in conjunction with the underlying large scale structure distribution 
provided by the simulations.

In brief,
to produce a realistic synthetic AGN population: 
\begin{itemize}
\item we used the SMBH list of the cosmo-OWLS simulations \citep[][AGN8.0 feedback model, WMAP7 cosmology]{Lebrun14},
\item we used the empirical assessment of the bolometric corrections by \citet{Lusso12} to convert the simulated AGN bolometric luminosities to X-ray emission,
\item we applied the obscuration function by \citet{Ueda14} to compute the column density of the AGN torus and the observed X-ray flux, 
\item we modeled the X-ray background by adding (a) the X-ray photon and solar proton contribution following Snowden et al. (2008), 
and (b) the particle background from 200 ks closed filter wheel {\it XMM-Newton} exposures, and 
\item we simulated all instrumental and survey-dependent signatures.
\end{itemize}

We argue that the described methodology can be applied on
the output of next generation hydrodynamical simulations \citep[e.g.
BAHAMAS:][]{Mccarthy17}, while, by adjusting the instrumental and the survey-dependent parameters,
the produced synthetic AGN catalogs can provide predictions for
future X-ray missions.

\acknowledgements
We would like to thank the anonymous referee for constructive comments that have helped us to improve the quality of this paper. We would like to thank James Aird, Johannes Buchner and Yetli Rosas-Guevara for providing their data and Joop Schaye for helpful discussions. XXL is an international project based around an {\it XMM-Newton} 
Very Large Programme surveying two 25 $deg^2$
extragalactic fields at a depth of $5\times10^{-15}$ erg s$^{-1}$ cm$^{-2}$ in [0.5-2] keV at the 90\% completeness level (see XXL paper I).
The XXL website is
http://irfu.cea.fr/xxl. Multiband information and spectroscopic follow-up of the
X-ray sources are obtained through a number of survey programmes, summarized at http://xxlmultiwave.pbworks.com/.
EK acknowledges the Centre National d’Etudes Spatiales 
(CNES) and CNRS for support of post-doctoral research. FP acknowledges support by the German Aerospace Agency (DLR) with funds
from the Ministry of Economy and Technology (BMWi) through grant 50 OR 1514 and grant 50 OR 1608. 
\bibliographystyle{aa}
\bibliography{mylib}

\end{document}